\documentclass[aps,reprint,longbibliography,amsmath,amssymb,prb]{revtex4-2} 
\usepackage{graphicx}
\usepackage{dcolumn}
\usepackage{bm}
\usepackage{mathtools}
\usepackage{dsfont}
\usepackage{nicematrix}
\usepackage{braket} 
\usepackage{silence} 
\usepackage[colorlinks=true, linkcolor=blue, citecolor=red]{hyperref}
\usepackage{cleveref} 

\begin{document}

\title{Quantum geometry of the rotating shallow water model}

\author{Sriram Ganeshan}
\affiliation{Department of Physics, City College, City University of New York, New York, NY 10031}
\affiliation{CUNY Graduate Center, New York, NY 10031}
\author{Alan T. Dorsey}
\affiliation{Department of Physics and Astronomy, University of Georgia, Athens, Georgia 30602}

\date{\today} 

\begin{abstract}

The rotating shallow water equations (RSWE) are a mainstay of atmospheric and oceanic modeling, and their wave dynamics has close analogues in settings ranging from two-dimensional electron gases to active-matter fluids. While recent work has emphasized the topological character of RSWE wave bands, here we develop a complementary quantum-geometric description by computing the full quantum geometric tensor (QGT) for the linearized RSWE on an $f$-plane. The QGT unifies two pieces of band geometry: its real part defines a metric that quantifies how rapidly wave polarization changes with parameters, while its imaginary part is the Berry curvature that controls geometric phases and topological invariants. We obtain compact, symmetry-guided expressions for all three bands, highlighting the transverse structure of the metric and the monopole-like Berry curvature that yields Chern numbers for the Poincaré bands. Finally, we describe a feasible route to probing this geometry in rotating-tank experiments via weak, time-periodic parametric driving.

\end{abstract}
	
\maketitle

\section{Introduction}

The rotating shallow water equations (RSWE) form the foundation for the modeling of atmospheric and oceanic fluids, providing a minimal yet remarkably rich description of large-scale flows and wave phenomena in a thin fluid layer under rotation \cite{vallis2017atmospheric,zeitlin2018geophysical}. In the familiar $f$-plane approximation, the RSWE captures the interplay of stratification and Coriolis forces that organizes motion into fast Poincar\'e (inertia--gravity) waves and a slow geostrophic mode. Beyond geophysics, closely related linear wave problems arise in diverse settings, including two-dimensional electron fluids in a perpendicular magnetic field (magnetoplasmons) \cite{jin2016topological,finnigan2022equatorial}, active or chiral fluids with parity- and time-reversal-breaking stresses \cite{souslov2019topological}, and other continuum media \cite{parker2020topological} in which rotation- or Lorentz-like couplings endow wavebands with nontrivial geometric structure.

A key recent development is the realization that linear wave equations in continuous media can exhibit the same geometric and topological phenomena familiar from Bloch bands in condensed matter physics \cite{bernevig2013topological,vanderbilt2018berry}. In particular, the seminal work of Delplace \textit{et al}.~\cite{delplace2017topological} showed that after the RSWE are linearized and written as a Hermitian eigenvalue problem, the resulting three-band structure supports Berry curvature and quantized Chern numbers for the Poincar\'e bands. This viewpoint has clarified the robustness of equatorial and coastal wave guiding as a \textit{bulk--boundary correspondence} in a continuum model, and it has connected geophysical wave dynamics to the broader theme of topological phases \cite{delplace2017topological, tauber2019bulk, delplace2022berry,perez2021manifestation, venaille2023ray}. 
While much of the existing literature focuses on topology—i.e., Berry curvature integrated over parameter space—many physical consequences of band geometry are not purely topological and cannot be inferred from Berry curvature alone.

In condensed matter systems, the natural framework that unifies these ideas is the \textit{quantum geometric tensor} (QGT) of an isolated band. For a normalized eigenmode $|u_n(\bm{\lambda})\rangle$ depending smoothly on adiabatic control parameters $\bm{\lambda}=(\lambda^1,\lambda^2,\dots)$, the QGT is the gauge-invariant Hermitian tensor
that measures the overlap of projected tangent vectors in Hilbert space. Its real, symmetric part is the Fubini--Study (FS) metric $g_{ij}$, which quantifies the infinitesimal distinguishability of nearby eigenstates and endows parameter space with a Riemannian structure; its imaginary, antisymmetric part is the Berry curvature $F_{ij}$, which defines a symplectic two-form and controls geometric phases. In condensed matter physics, the FS metric has become increasingly important: it constrains response functions, enters bounds and inequalities linking geometry and topology, and plays a direct dynamical role in flat and nearly flat bands; for example, through the geometric contribution to superfluid weight and related notions of ``quantum distance'' \cite{peotta2015superfluidity,kolodrubetz2017geometry,ozawa2018extracting}. These developments motivate a parallel program in the classical waves of geophysical fluid dynamics: once the RSWE are viewed as a multiband wave system, the full band geometry, and not just the integrated Berry curvature, should control measurable aspects of wave propagation and forcing. We stress that the notion of the quantum geometric tensor originates in condensed-matter physics, where it is formulated for genuinely quantum systems. Nevertheless, an analogous construction exists for the linearized classical wave modes of the RSWE, just as Berry curvature can be meaningfully defined for classical continuum waves. 

In this paper, we develop a quantum geometric-like description of the RSWE by computing the full QGT for all three bands in the $f$-plane approximation. 
We highlight how symmetry fixes much of its structure: rotational covariance in the $(k_x,k_y)$ plane leads to a transverse metric with a null direction associated with the radial parameter, while the Berry curvature takes the form of an effective monopole field in parameter space for the Poincar\'e bands. We then show how these geometric quantities can be operationally accessed 
using the general idea of geometric spectroscopy, in which periodic modulation of control parameters produces interband transitions whose integrated rates isolate components of $g_{ij}$ and $F_{ij}$ \cite{ozawa2018extracting,ozawa2019probing}. 

The remainder of the paper is organized as follows. In Sec.~\ref{sec:RSWE} we review the linearized RSWE on the $f$-plane and formulate the problem as a three-band Hermitian eigenvalue problem. In Sec.~\ref{sec:eigenvalues} we exploit the pseudospin-1 structure of the linearized dynamics to obtain eigenfrequencies, eigenvectors, and a transparent helicity interpretation of the Poincar\'e and geostrophic modes. In Sec.~\ref{sec:QGT} we compute the QGT, summarize its symmetry-determined tensor structure, and derive band-dependent geometric and topological invariants. In Sec.~\ref{sec:measurement} we discuss measurement protocols based on time-dependent parametric forcing and connect integrated transition rates to the FS metric and curvature. We conclude in Sec.~\ref{sec:summary} with a summary and outlook, including prospects for extending the framework to study geophysical ray-tracing in spatially varying backgrounds and weakly nonlinear regimes.

\section{The Rotating Shallow Water Equations}\label{sec:RSWE}

\begin{figure}
    \centering
    \includegraphics[width=1.0\linewidth]{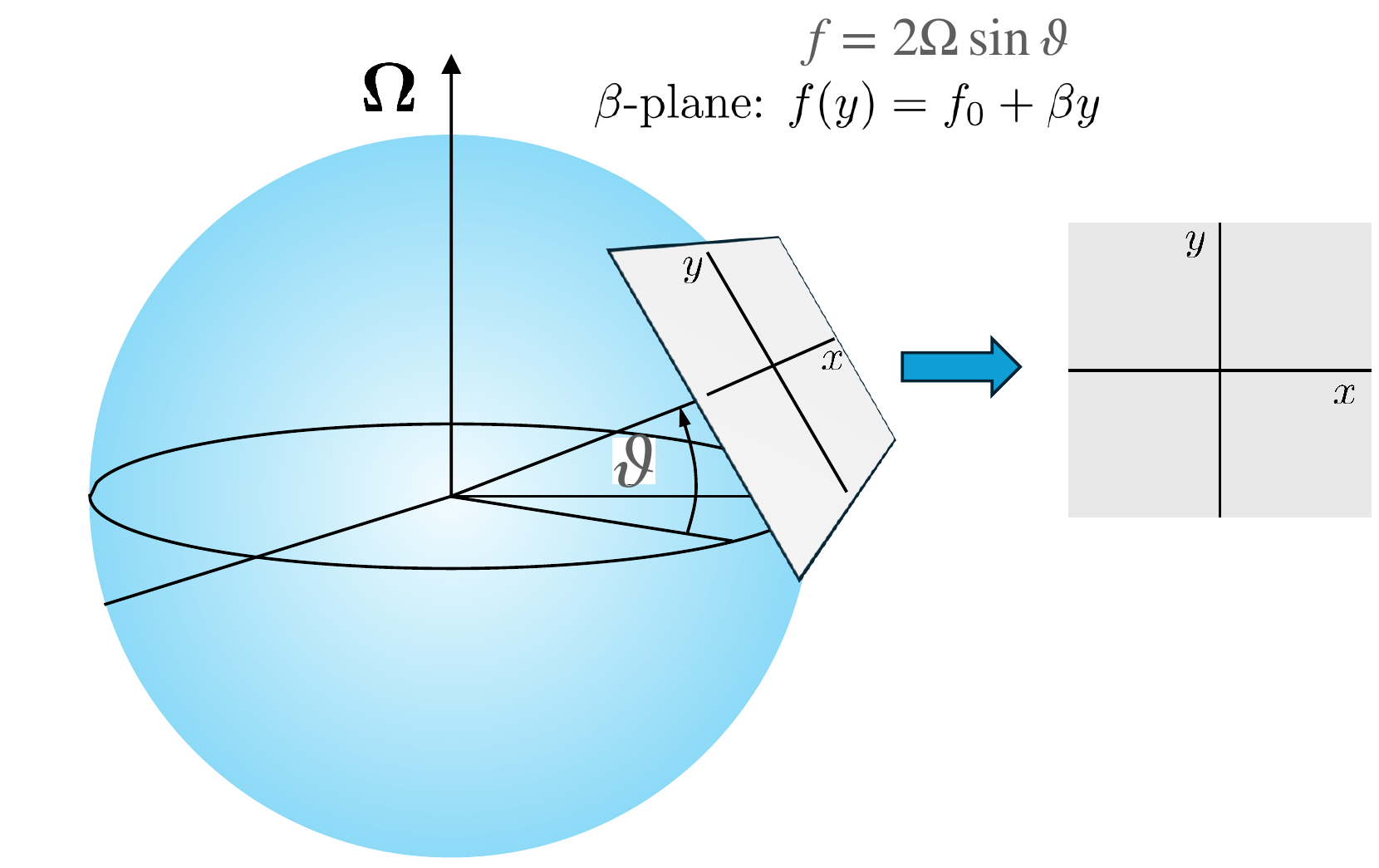}
    \caption{The $\beta$-plane on a rotating planet. The angular velocity $\boldsymbol{\Omega}$ is aligned with the rotation axis; the tangent plane is at a latitude $\vartheta$ with respect to the equatorial plane; the Coriolis parameter is $f=2\Omega \sin\vartheta$, which becomes $f=f_0+\beta y$ in the $\beta$-plane approximation; and $(x,y)$ are the local Cartesian coordinates on the tangent plane. }
    \label{fig:beta-plane}
\end{figure}

The rotating shallow water equations (RSWE) \cite{zeitlin2018geophysical,vallis2017atmospheric} for a thin layer of fluid of height $h(\mathbf{x},t)$ and velocity $\mathbf{u}$ on a two-dimensional surface are
\begin{align}
	\partial_t h +\nabla \cdot \left (h\mathbf{u}\right) &= 0,\\
	\partial_t \mathbf{u} + (\mathbf{u} \cdot \mathbf{\nabla}) \mathbf{u} &= -g \mathbf{\nabla} h - f \hat{\mathbf{n}} \times \mathbf{u}.
	\label{eq:momentum}
\end{align}
Here $g$ is the constant of gravitational acceleration; $f = 2\boldsymbol{\Omega} \cdot \hat{\mathbf{n}} = 2 \Omega \sin \vartheta$ is the \textit{Coriolis parameter}; $\boldsymbol{\Omega}= \Omega \hat{\mathbf{z}}$ is the Earth's angular velocity about the axis of rotation $z$; $\hat{\mathbf{n}}$ is the local unit normal to the Earth's surface; and $\vartheta$ is the latitude measured from the equator [see Fig.~\eqref{fig:beta-plane} for the coordinates and notation; we use the script $\vartheta$ to denote the latitude and distinguish it from the polar angle $\theta$ used later in this work]. The Coriolis parameter takes its maximum values of $f=\pm 2\Omega$ at the North and South poles and vanishes along the equator. 

To facilitate and simplify calculations one works on a tangent-plane affixed to the Earth's surface at a latitude $\vartheta_0$, with local Cartesian coordinates $(x,y)$, where $x$ is the zonal or east-west coordinate and $y$ the meridional or north-south coordinate. The \textit{$f$-plane approximation} replaces the Coriolis parameter with a constant, $f_0=2\Omega \sin\vartheta_0$, with $f_0>0$ in the Northern Hemisphere and $f_0<0$ in the Southern Hemisphere. The \textit{$\beta$-plane approximation} retains the first derivative of the Coriolis parameter in the meridional coordinate $y$ so that $f(y) = f_0 + \beta y$, where $\beta=2\Omega\cos\vartheta_0/R$ (with $R$ the Earth's radius) is the \textit{Rossby parameter}. The Rossby parameter is important in determining the dynamics of Rossby waves, as well as wave phenomena near the equator, where $f_0=0$. By reducing the complexity of the full spherical system, the $\beta$-plane approximation provides an analytically tractable model that captures the essential physics of equatorial fluid dynamics while serving as a foundation for more complex numerical simulations. 

The linearized rotating shallow water equations are 
\begin{align}
	\partial_t u &= - g H \partial_x \eta + f v, \label{eq:rsw11} \\
	\partial_t v &= -gH \partial_y \eta -f u,  \label{eq:rsw12} \\
	\partial_t \eta &= -  (\partial_x u +\partial_y v). \label{eq:rsw13}
\end{align}
Here ${\bf v}=(u,v)$ are the $(x,y)$ components of the two-dimensional velocity and $\eta$ is the dimensionless relative height fluctuation defined by $\eta({\bf x},t)= [h ({\bf x},t) - H]/H$ with $H$ the mean height. For a nonrotating fluid ($f=0$) this system of equations has wave solutions, the \textit{shallow water gravity waves} of speed $c=\sqrt{gH}$. With this in mind, we nondimensionalize Eqs.~\eqref{eq:rsw11}-\eqref{eq:rsw13} by rescaling the velocities by $c$; time by $1/(2\Omega)$; and lengths by the Rossby deformation radius $L_d=c/(2\Omega$). After these rescalings, in terms of the dimensionless variables (labeled with tildes) we obtain
\begin{align}
	\partial_{\tilde{t}} \tilde{u} &= - \partial_{\tilde{x}} \eta + \tilde{f} \tilde{v},\label{eq:rsw21}\\
	\partial_{\tilde{t}} \tilde{v} &= -\partial_{\tilde{y}} \eta -\tilde{f} \tilde{u},\label{eq:rsw22}\\
	\partial_{\tilde t} \eta &= - \partial_{\tilde{x}} \tilde{u} - \partial_{\tilde{y}} \tilde{v} \label{eq:rsw23}.
\end{align}
For notational clarity we will drop the tildes in what follows. 

The three fields can be combined into a single three component state vector $|\Psi\rangle = (u, v, \eta)^{\mathsf{T}}$ (with $\mathsf{T}$ denoting the transpose), and the above linear system of equations may be written in the form of a Schr\"odinger equation as 
\begin{equation}\label{eq:hamiltonian}
i \partial_t |\Psi\rangle = {\cal H}_\mathrm{SWE} | \Psi \rangle,
\end{equation}
where ${\cal H}_\text{SWE}$ plays the role of a Hermitian Hamiltonian,
\begin{equation}\label{eq:hamiltonian1}
 \quad {\cal H}_\text{SWE} =
\begin{pmatrix}
	0 & i f & -i\partial_x \\
	-i f & 0 & -i\partial_{y} \\
	-i\partial_{x} & -i\partial_{y} & 0
\end{pmatrix}
.
\end{equation}
 The translational symmetry ensures the eigenmodes are proportional to $e^{i\left(k_x x + k_y y - \omega t\right)}$, so we may make the replacements $i\partial_t \rightarrow \omega$, $-i\partial_x \rightarrow k_x$, and $-i\partial_y \rightarrow k_y$, leading to the eigenvalue problem  
\begin{equation} \label{eq:hamiltonian2}
\omega |\Psi\rangle = {\cal H}_\text{SWE} |\Psi\rangle, \quad {\cal H}_\text{SWE}
  = 
  \begin{pmatrix}
      0 & if & k_x \\
      -if & 0  & k_y \\
      k_x & k_y & 0
  \end{pmatrix}
  .
\end{equation}

\section{Eigenvalues and eigenfunctions}\label{sec:eigenvalues}

\subsection{Reduction to a pseudospin-1 Weyl fermion Hamiltonian}

While Eq.~\eqref{eq:hamiltonian2} is the standard form for the linearized RSWE, it is useful in our formal development to rewrite Eq.~\eqref{eq:hamiltonian2} in a different representation using a basis $ |\chi\rangle  = U |\Psi\rangle = (v, -u, i\eta)^{\mathsf{T}}$, where the unitary operator $U$ is 
\begin{equation}\label{eq:unitary}
U=\left(\begin{NiceArray}{rrr}
0 & 1 & 0 \\
-1 & 0 & 0 \\
0 & 0 & i
\end{NiceArray}\right), \quad 
U^{-1}=U^{\dagger} = \left(\begin{NiceArray}{rrr}
0 & -1 & 0 \\
1 & 0 & 0 \\
0 & 0 & -i
\end{NiceArray}\right).
\end{equation}
The eigenfunctions in this new basis will be determined by $\omega|\chi\rangle = {\cal H}_\chi |\chi \rangle$, with
\begin{equation}\label{eq:hamiltonian3}
{\cal H}_\chi = U {\cal H}_\text{SWE} U^{-1} =
 \begin{pmatrix}
	0 & if & -ik_y \\
	- if & 0 & ik_x \\
	ik_y & -ik_x & 0
\end{pmatrix}
.
\end{equation}
We can rewrite this as 
\begin{equation}\label{eq:hamiltonian4}
    {\cal H}_\chi = -\mathbf{d}\cdot \mathbf{S}, \quad \mathbf{d} = (k_x,k_y,f),
\end{equation}
where the matrices $S_i$ are the Hermitian generators of the three-dimensional rotation group $SO(3)$ in the spin-1 representation, given by $\left(S_i\right)_{j k}=-i \epsilon_{i j k}$ with $[S_i,S_j]=i\epsilon_{ijk}S_k$, where $\epsilon_{ijk}$ is the fully antisymmetric Levi-Civita tensor. 

Equation \eqref{eq:hamiltonian4} is the low-energy effective Hamiltonian for a threefold linear band crossing, which describes a spin-1 generalization of a Weyl fermion (or equivalently, a triple-point fermion), studied extensively in the context of topological materials \cite{bradlyn2016beyond,mai2018topological,winkler2019topology}. Variants of Eq.~\eqref{eq:hamiltonian4} have appeared in treatments of topological waves in geophysical fluids: for instance, see Delplace \cite{delplace2022berry} and Perez \cite{perez2022topological}. 

In what follows we will need to identify our adiabatic control parameters---the parameters that we use to characterize the space of solutions to Eq.~\eqref{eq:hamiltonian}. In treatments of band topology and geometry in condensed matter systems \cite{bernevig2013topological,vanderbilt2018berry} these control parameters would be the crystal momenta in the Brillouin zone $(k_x,k_y)$ (for a two-dimensional material); in our treatment we include the wavenumbers $(k_x,k_y)$ as well as the Coriolis parameter $f$ as our adiabatic control parameters, with $f$ constituting the $z$-component of the control vector $\mathbf{d}=(k_x,k_y,f)$. Including $f$ as one of the control parameters lends some mathematical simplicity to the problem. And while changing the Coriolis parameter in a geophysics setting might seem unphysical (or at least difficult), it would be natural in analogue systems such a two-dimensional electron gas in a perpendicular magnetic field, where the magnetic field  plays the role of the Coriolis parameter. Later in the paper we will explicitly identify components of the QGT that one would obtain if the Coriolis parameter were held constant. 

\subsection{Calculation of the eigenvalues and eigenfunctions}

In the spin-1 representation of $SO(3)$, the action of the Hamiltonian ${\cal H}_\chi= -\mathbf{d} \cdot \mathbf{S}$ on a vector $\mathbf{a}$ is ${\cal H}_\chi \mathbf{a} = -i \mathbf{d} \times \mathbf{a}$, 
so ${\cal H}_\chi$ generates rotations of $\mathbf{a}$ about the unit direction vector
 $\hat{ \mathbf{d}} =  \mathbf{d}/d $, where $d=|\mathbf{d}|=\sqrt{k_x^2+k_y^2+f^2}$. 
We immediately see that one eigenfunction of ${\cal H}_\chi$ is $|u_0\rangle = \hat{\mathbf{d}} = \mathbf{e}_r$ with eigenvalue $\omega_0=0$; this is the dispersionless ``flat band'' of geostrophic balance, where the Coriolis force is balanced by pressure gradients. The remaining two orthonormal eigenfunctions determine a plane normal to $\mathbf{e}_r$; this plane will be defined by the polar unit vector $\mathbf{e}_\theta$ and the azimuthal unit $\mathbf{e}_\phi$, so that $(\mathbf{e}_r,\mathbf{e}_\theta, \mathbf{e}_\phi)$ form an orthogonal right-handed triad. Introducing the \textit{circularly polarized states} (note the sign convention!) 
\begin{equation}\label{eq:polarized}
    \mathbf{e}_{ \pm}= \frac{\mathbf{e}_\theta \mp i \mathbf{e}_\phi}{\sqrt{2}},
\end{equation}
and using $\mathbf{d} \times \mathbf{e}_\theta= d\, \mathbf{e}_\phi$, $\mathbf{d} \times \mathbf{e}_\phi=-d\, \mathbf{e}_\theta$,
we obtain
\begin{equation}
    {\cal H}_\chi \mathbf{e}_{\pm} = -i \mathbf{d}\times \mathbf{e}_\pm = \pm d\,\mathbf{e}_\pm ,
\end{equation}
so the remaining two eigenfunctions are $\mathbf{e}_\pm$ with eigenvalues $\omega_\pm = \pm d$. These two bands are the Poincar\'e waves (sometimes called inertia-gravity waves or IGW).  
Introducing the standard spherical angles $(\theta,\phi)$, 
\begin{align}
 \mathbf{e}_r &=(\sin \theta \cos \phi, \sin \theta \sin \phi, \cos \theta)^\mathsf{T}, \\
\mathbf{e}_\theta &=(\cos \theta \cos \phi, \cos \theta \sin \phi,-\sin \theta)^\mathsf{T}, \\
\quad \mathbf{e}_\phi &=(-\sin \phi, \cos \phi, 0)^\mathsf{T}, 
\end{align}
the normalized eigenfunctions, with eigenvalues $\omega_m = m d $, $m \in \{-1,0,1\}$ are
\begin{align}
\left|u_0\right\rangle & =\mathbf{e}_r 
 = \begin{pmatrix} \sin\theta\cos\phi \\ \sin\theta\cos\phi \\ \cos\theta \end{pmatrix} ,
 \\
\left|u_{ \pm}\right\rangle &=\frac{1}{\sqrt{2}}\left(\mathbf{e}_\theta \mp i \mathbf{e}_\phi\right) 
 = \frac{1}{\sqrt{2}} \begin{pmatrix} \cos\theta\cos\phi \pm i \sin\phi\\ \cos\theta\sin\phi \mp i \cos\phi\\ - \sin\theta \end{pmatrix}.
\end{align}
\textit{Notice how the eigenvectors only depend on the angles $(\theta,\phi)$, and not on $d$, while the eigenvalues only depend on $d$ and not on $(\theta,\phi)$}. This is a consequence of the rotational covariance of our Hamiltonian ${\cal H}_\chi$, and will have important implications for the structure of the QGT, as we will see in the next section. 

While the spherical coordinates most conveniently capture the rotational symmetry of the eigenfunctions, for some purposes we might prefer the Cartesian coordinates $(k_x,k_y,f)$; introducing $k_{\perp}=\sqrt{k_x^2+k_y^2}$, we have 
\begin{equation}
\sin \theta=\frac{k_\perp}{d}, \quad \cos \theta=\frac{f}{d}, \quad \sin \phi=\frac{k_y}{k_\perp}, \quad\cos \phi=\frac{k_x}{k_\perp}, 
  \label{eq:def-kperp}
\end{equation}
so the eigenvectors become
\begin{equation}
      \lvert u_{0}(\mathbf k)\rangle =
      \frac{1}{d}
      \begin{pmatrix}
        k_x\\
        k_y\\
        f
      \end{pmatrix},
\end{equation}
\begin{equation}
      \lvert u_{\pm}(\mathbf k)\rangle =
      \frac{1}{\sqrt{2}}
      \begin{pmatrix}
        \dfrac{f k_x}{d\,k_\perp} \pm i\,\dfrac{k_y}{k_\perp}
        \\
        \dfrac{f k_y}{d\,k_\perp} \mp i\,\dfrac{k_x}{k_\perp}
        \\
        -\,\dfrac{k_\perp}{d}
      \end{pmatrix}.
\end{equation}

The eigenvectors are also states of definite \textit{helicity}, defined here as the projection of the pseudospin $\mathbf{S}$ (which acts as an internal angular momentum) along the direction $\hat{\mathbf{d}}$:
\begin{equation}
    \hat{h} \coloneq  \hat{\mathbf{d}}\cdot \mathbf{S}. 
\end{equation}
The geostrophic mode has $ \hat{h} |u_0\rangle = (0) |u_0\rangle$, so it has helicity 0; the two Poincar\'e modes have $\hat{h} |u_\pm\rangle = (\pm 1) |u_\pm \rangle$, so the upper Poincar\'e mode has helicity $+1$ and the lower mode has helicity $-1$. The helicity captures the state of polarization of the modes---the Poincar\'e modes are either left or right circularly polarized about the radial vector $\hat{\mathbf{d}}$ [as is evident from the eigenstates $\mathbf{e}_\pm$ in Eq.~\eqref{eq:polarized}], while the geostrophic mode is longitudinal in $\mathbf{d}$-space. The results for the eigenvalues and eigenvectors are summarized in Table~\ref{tab:RSWE-bands}.

\begin{table*}[t]
  \centering
 \small
  \setlength{\tabcolsep}{3pt}
  \renewcommand{\arraystretch}{1.3}
  \begin{tabular}{c c c c}
    \hline
    \shortstack{Band $m$ \\[2pt] $\omega_m = m d$} &
    \shortstack{Polarization \&\\helicity} &
    \shortstack{Cartesian eigenvectors\\$\lvert u_m(\mathbf{k})\rangle$} &
    \shortstack{Spherical eigenvectors\\$\lvert u_m(\theta,\phi)\rangle$}
    \\
    \hline\hline
    \shortstack{$m=\pm1$ \\[2pt] Poincar\'e}
    &
    \shortstack{$\lvert u_\pm\rangle\perp\mathbf d$ \\[2pt] Transverse \\[2pt] Helicity $\pm1$}
    &
    $\displaystyle
      \lvert u_{\pm}(\mathbf k)\rangle =
      \frac{1}{\sqrt{2}}
      \begin{pmatrix}
        \dfrac{f k_x}{d\,k_\perp} \pm i\,\dfrac{k_y}{k_\perp}
        \\[3pt]
        \dfrac{f k_y}{d\,k_\perp} \mp i\,\dfrac{k_x}{k_\perp}
        \\[2pt]
        -\,\dfrac{k_\perp}{d}
      \end{pmatrix}
      $
    &
    $\displaystyle
      \lvert u_{\pm}(\theta,\phi)\rangle =
      \frac{1}{\sqrt{2}}
      \begin{pmatrix}
        \cos\theta\cos\phi \pm i\,\sin\phi
        \\[2pt]
        \cos\theta\sin\phi \mp i\,\cos\phi
        \\[2pt]
        -\,\sin\theta
      \end{pmatrix}
      $
    \\[4pt]
    \hline
    \shortstack{$m=0$ \\[2pt] Geostrophic}
    &
    \shortstack{$\lvert u_0\rangle\parallel\mathbf d$ \\[2pt] Longitudinal \\[2pt] Helicity $0$}
    &
    $\displaystyle
      \lvert u_{0}(\mathbf k)\rangle =
      \frac{1}{d}
      \begin{pmatrix}
        k_x\\[2pt]
        k_y\\[2pt]
        f
      \end{pmatrix}
      $
    &
    $\displaystyle
      \lvert u_{0}(\theta,\phi)\rangle =
      \begin{pmatrix}
        \sin\theta\cos\phi
        \\[2pt]
        \sin\theta\sin\phi
        \\[2pt]
        \cos\theta
      \end{pmatrix}
      $
    \\[4pt]
    \hline
  \end{tabular}
  \caption{Normalized eigenvectors of the pseudospin-1 rotating shallow water (RSWE) Hamiltonian
  ${\cal H}_\chi(\mathbf k) = -\mathbf d\cdot\mathbf S$ [Eq.~\eqref{eq:hamiltonian4}] with $\mathbf d=(k_x,k_y,f)$ and
  magnitude $d = |\mathbf d|$. The dynamical variables are
  $(u,v,\eta)$ (zonal velocity, meridional velocity, free-surface displacement).
  The Poincar\'e bands $m=\pm1$ are transverse and circularly polarized (helicity $\pm1$),
  while the $m=0$ band is geostrophic and longitudinal along $\mathbf d$ (helicity $0$).
  The horizontal wavenumber $k_\perp$ and angles $(\theta,\phi)$ are defined in
  Eq.~\eqref{eq:def-kperp}. The frequencies satisfy $\omega_m = m d$.}
  \label{tab:RSWE-bands}
\end{table*}

\section{Calculation of the quantum geometric tensor}\label{sec:QGT}

With the eigenvalues and eigenvectors in hand, we are in a position to calculate the components of the QGT. The properties of the QGT are summarized in the Appendix \ref{app:overview}, and we refer the reader to this appendix as well as to several review articles \cite{cheng2013quantum,verma2025quantum,gao2025quantum,yu2024quantum} and the original literature \cite{provost1980riemannian, berry1989quantum, fubini1904sulle, study1905kurzeste} for an introduction to the QGT and its applications. 

Suppose $\bm{\lambda}=(\lambda^1,\ldots,\lambda^N)$ forms a set of $N$ real control parameters and let $|u_m(\bm{\lambda})\rangle$ be a \emph{normalized} eigenvector of the $m$-th nondegenerate band  of a Hermitian Hamiltonian ${\cal H}(\bm{\lambda})$. The \emph{quantum geometric tensor} (QGT) is the gauge-invariant, complex Hermitian tensor
\begin{align}
Q_{ij}^{(m)}(\bm{\lambda}) & \coloneq \big\langle \partial_i u_m \,\big|\,\bigl( \mathds{1}_N - |u_m\rangle\langle u_m| \bigr)\,\big|\, \partial_j u_m \big\rangle \nonumber\\
&= g_{ij}^{(m)} + \frac{i}{2} F_{ij}^{(m)},
\label{eq:QGT1}
\end{align}
where $\partial_i \coloneq \partial/\partial\lambda^i$ and 
where the real part $g^{(m)}_{ij}$ is the symmetric \textit{quantum metric tensor} [or Fubini-Study (FS) metric] for the $m$-th band and the imaginary part $F^{(m)}_{ij}$ is the antisymmetric \textit{Berry curvature} for the $m$-th band. For our RSWE, our control (adiabatic) parameters will be $(k_x,k_y,f)$. For practical purposes using Eq.~\eqref{eq:QGT1} directly to calculate the QGT can be cumbersome, as it requires taking derivatives of the \textit{normalized} eigenvectors with respect to the control parameters (the normalization will generally depend upon the control parameters). As an alternative, we will calculate the QGT using the spectral representation in Eq.~\eqref{eq:spectral}, which has the added benefit of providing a straightforward generalization of our results for our pseudospin-1 model to higher spin-$S$. 

The QGT for the generalized pseudospin-$S$ model ${\cal H}_\chi=-{\bf d}\cdot {\bf S}$ on a nondegenerate band with parameters $(d_x,d_y,d_z)$ and the spin algebra $[S_i,S_j]=i\epsilon_{ijk} S_k$, is
\begin{equation} \label{eq:spectral_ab}
    Q_{ij}^{(m)} =\sum_{n \neq m} \frac{\langle u_m| \partial_{i} {\cal H}_\chi|u_n\rangle\langle u_n| \partial_{j} {\cal H}_\chi|u_m\rangle}{\left(\omega_n-\omega_m\right)^2},
\end{equation}
with $i,j\in (x,y,z)$. The denominator in Eq.~\eqref{eq:spectral_ab} is $\left(\omega_n-\omega_m\right)^2=d^2(n-m)^2$; combining this with $\partial_{i} {\cal H}_\chi = \partial {\cal H}_\chi/\partial d_i = -S_i$,  Eq.~\eqref{eq:spectral_ab} becomes
\begin{equation}\label{eq:spectral_ab1}
Q_{ij}^{(m)}= \frac{1}{d^2}\sum_{n \neq m} \frac{\langle u_m| S_i|u_n\rangle\langle u_n| S_j|u_m\rangle}{\left(n-m\right)^2} .
\end{equation}

To evaluate Eq.~\eqref{eq:spectral_ab1}, we work in a basis that diagonalizes $S_z$ and $S^2$, the states
$|u_m\rangle=|S,m\rangle$, with total spin $S$ and $m\in (-S...S)$. We start with components of Eq.~\eqref{eq:spectral_ab1} with either $i$ or $j$ equal to $z$. Since $S_z$ is diagonal in the $|S,m\rangle$ basis, $\langle S,m| S_z|S,n\rangle=m \delta_{m n}$,
and every term in the sum in Eq.~\eqref{eq:spectral_ab} with $n \neq m$ vanishes, so that components of the QGT with either $i$ or $j$ equal to $z$ vanish:
\begin{equation}
Q_{x z}^{(m)} = Q_{y z}^{(m)} = Q_{z x}^{(m)} = Q_{z y}^{(m)}= Q_{z z}^{(m)} =0 . 
\end{equation}
To calculate the remaining four components, we introduce the raising and lowering operators for spin $\mathbf{S}$, $S_{ \pm}=S_x \pm i S_y$. Matrix elements of these operators are only nonzero for $n=m \pm 1$ (for a discussion, see Sec.~27 of Ref.~\cite{landau2013quantum}); expressing $S_x$ and $S_y$ in terms of $S_\pm$, we find the only nonzero matrix elements are [we use the shorthand notation of $|m\rangle$ for $|S,m\rangle$] 
\begin{align}
\langle m| S_x|m+1\rangle & =\frac{1}{2} \beta_m, & \langle m+1| S_y|m\rangle & =\frac{1}{2 i} \beta_m, \\
\langle m| S_x|m-1\rangle & =\frac{1}{2} \alpha_m, & \langle m-1| S_y|m\rangle & =-\frac{1}{2 i} \alpha_m,
\end{align}
where the real constants $\alpha_m$ and $\beta_m$ are given by \cite{landau2013quantum}
 \begin{align}
\alpha_m &= \langle m| S_{+}|m-1\rangle=\sqrt{S(S+1)-m(m-1)}, \\
\beta_m &= \langle m| S_{-}|m+1\rangle=\sqrt{S(S+1)-m(m+1)} .
\end{align}   
For $n=m \pm 1$ the denominator in Eq.~\eqref{eq:spectral_ab1}  is $(n-m)^2=1$, so Eq.~\eqref{eq:spectral_ab1} becomes
\begin{align} \label{eq:Qxx}
Q_{xx}^{(m)} = Q_{yy}^{(m)}  & =\frac{1}{d^2} \bigl[ |\langle m| S_x | m+1\rangle|^2 \nonumber \\
  & \qquad \qquad + |\langle m| S_x | m-1 \rangle |^2 \bigr] \nonumber \\
  & =\frac{1}{4d^2} ( \alpha_m^2 + \beta_m^2 ) = \frac{C_m}{d^2},
\end{align}
where
\begin{equation}\label{eq:Cm}
    C_m = \frac{S(S+1)-m^2}{2} \overset{S=1}\Longrightarrow C_m=\frac{2-m^2}{2}.
\end{equation}
This result for $C_m$ for general $S$ agrees with the previous results obtained using different methods \cite{lin2021dual,graf2021berry}. For the off-diagonal $xy$-component we have
\begin{align}\label{eq:Qxy}
    Q_{xy}^{(m)} = -Q_{yx}^{(m)} &= \frac{1}{d^2} \bigl[ \langle m | S_x | m+1 \rangle \langle m+1 | S_y | m\rangle \nonumber \\
     & \qquad + \langle m | S_x | m-1 \rangle \langle m-1 | S_y | m\rangle \bigr] \nonumber \\
     & = \frac{i}{4 d^2} \left[ \alpha_m^2 - \beta_m^2 \right] = \frac{i m }{2d^2}.
\end{align}
Recalling that $g_{ij}^{(m)} = \mathrm{Re}\, Q_{ij}^{(m)}$ and $F_{ij}^{(m)} = 2\, \mathrm{Im}\, Q_{ij}^{(m)}$, we can combine Eqs.~\eqref{eq:Qxx} and \eqref{eq:Qxy} into the following compact forms [with $\hat{\mathbf{z}}^{\mathsf{T}} = (0,0,1)$]:
\begin{equation}\label{eq:metric_z}
g_{ij}^{(m)}(\hat{\mathbf{z}})=\frac{C_m}{ d^2}\left(\mathds{1}_3 -\hat{\mathbf{z}} \hat{\mathbf{z}}^{\mathsf{T}}\right)_{ij},
\end{equation}
\begin{equation}\label{eq:curvature_z}
    F_{ij}^{(m)} (\hat{\mathbf{z}})  = m \frac{\epsilon_{ijz}}{d^2} .
\end{equation}

We have succeeded in computing all of the components of the QGT, but only for a specific orientation $\hat {\mathbf {z}}$. As a final step, we want to rotate $\hat{\mathbf{z}}$ back to a general direction $\hat{\mathbf{d}}$. This is accomplished through a real rotation $R \in SO(3)$ such that $R \hat{\mathbf{z}}=\hat{\mathbf{d}}$. We will carry this out separately for the FS metric and for the Berry curvature. 

Since the FS metric $g_{ij}^{(m)}$ is a rank-2 tensor in $\mathbf{d}$-parameter space, under the rotation $R$ its components transform as
\begin{equation}
g^{(m)}(\hat{\mathbf{d}})=R g^{(m)}(\hat{\mathbf{z}}) R^{\mathsf{T}} .
\end{equation}
Using Eq.~\eqref{eq:metric_z},
\begin{align}
g^{(m)}(\hat{\mathbf{d}}) &=\frac{C_m}{ d^2} R\left(\mathds{1}_3 -\hat{\mathbf{z}} \hat{\mathbf{z}}^{\mathsf{T}}\right) R^{\mathsf{T}} \nonumber \\
 &=\frac{C_m}{ d^2}\left[\mathds{1}_3-(R \hat{\mathbf{z}})(R \hat{\mathbf{z}})^{\mathsf{T}}\right] \nonumber \\
 &=\frac{C_m}{ d^2}\left(I-\hat{\mathbf{d}} \hat{\mathbf{d}}^{\mathsf{T}}\right). 
\end{align}
In component form, this becomes
\begin{align} \label{eq:metric_tensor}
g_{ij}^{(m)}(\mathbf{d}) & =\frac{C_m}{d^2}\left(\delta_{ij}-\hat{d}_i \hat{d}_j\right) \nonumber\\
 & = C_m \left( \partial_i \hat{\mathbf{d}} \cdot \partial_j \hat{\mathbf{d}} \right),
\end{align}
where $\hat{d}_i = d_i/d$, $\mathbf{d} = (k_x,k_y, f)$ and $d=\sqrt{k_x^2 + k_y^2 + f^2}$; to derive the last line above, we have used $\partial_i \hat{d}_a=(\delta_{i a}-\hat{d}_a \hat{d}_i)/d$.
In an explicit component form with the Cartesian parameters $(k_x,k_y,f)$, 
\begin{equation}\label{eq:metric_matrix1}
g^{(m)}(k_x,k_y,f)=\frac{C_m}{d^4}
\left(
\begin{NiceArray}{cc:c}
k_y^2+f^2 & -k_x k_y & -k_x f\\
-k_x k_y & k_x^2+f^2 & -k_y f\\ \hdottedline
-k_x f & -k_y f & k_x^2+k_y^2
\end{NiceArray}
\right).
\end{equation}
The dotted lines in the matrix above indicate the upper-left $(k_x,k_y)$ block of the FS metric, in which the Coriolis parameter $f$ is held fixed. We can also calculate the FS metric in spherical coordinates $(k_x,k_y,f)=(d\sin\theta\cos\phi,\ d\sin\theta\sin\phi,\ d\cos\theta)$, with the simple result
\begin{equation}\label{eq:metric_matrix2}
g^{(m)}(\theta,\phi,d)=
\left(
\begin{NiceArray}{cc:c}
C_m & 0 & 0\\
0 & C_m\sin^2\theta & 0\\ \hdottedline
0 & 0 & 0
\end{NiceArray}
\right).
\quad
\end{equation}

Next, we work out the curvature tensor $F^{(m)}_{ij}$ for a general direction $\hat{\mathbf{d}}$. The natural generalization of Eq.~\eqref{eq:curvature_z} is
\begin{align}\label{eq:curvature1}
    F^{(m)}_{ij} &= m \frac{ \epsilon_{ijk} d_k }{d^3} \nonumber\\
     & = m\, \hat{\mathbf{d}} \cdot ( \partial_i \hat{\mathbf{d}} \times \partial_j \hat{\mathbf{d}} ),
\end{align}
where $m\in \{-1,0,1\}$ for the three bands. Equivalently, we can express this as a two-form (with $\lambda^i=d^i$)
\begin{equation}\label{eq:Berry_twoform}
F^{(m)} = \frac{1}{2}\,F_{ij}^{(m)}\,d\, \lambda^i\wedge d\, \lambda^j.
\end{equation}
It is again useful to display the curvature tensor in both Cartesian coordinates,
\begin{equation}\label{eq:curvature_matrix1}
F^{(m)}(k_x,k_y,f)=\frac{m}{d^3}
\left(
\begin{NiceArray}{cc:c}
0 & f & -k_y\\
-f & 0 & k_x\\ \hdottedline
k_y & -k_x & 0
\end{NiceArray}
\right),
\end{equation}
as well as spherical coordinates
\begin{equation}\label{eq:curvature_matrix2}
F^{(m)}(\theta,\phi,d)=
\left(
\begin{NiceArray}{cc:c}
0 & m\sin\theta & 0\\
-m\sin\theta & 0 & 0\\ \hdottedline
0 & 0 & 0
\end{NiceArray}
\right).
\end{equation}
Combining Eqs.~\eqref{eq:metric_tensor} and \eqref{eq:curvature1}, the full QGT expressed in terms of the normalized parameter vector $\hat{\mathbf{d}}$ can be written in the convenient and suggestive form
\begin{equation}\label{eq:QGT_final}
    Q_{ij}^{(m)} = C_m\, \partial_i \hat{\mathbf{d}}\cdot \partial_j \hat{\mathbf{d}}
     + i \frac{m}{2} \hat{\mathbf{d}}\cdot \left(\partial_i \hat{\mathbf{d}} \times \partial_j \hat{\mathbf{d}} \right).
\end{equation}
Equation \eqref{eq:QGT_final} is one of the central results of this paper. 

Having derived the full pseudospin-$S$ QGT, we would like to draw attention to several important features. 
\begin{itemize}

\item Both the FS metric and the Berry curvature have a \textit{null direction} along $\hat{\mathbf{d}}$, such that $g_{ij}^{(m)}\, \hat{d}_j = 0$ and  $F_{ij}^{(m)} \hat{d}_j = 0$. This is readily evident when using spherical coordinates---the block-diagonal forms of $g^{(m)}$ in Eq.~\eqref{eq:metric_matrix2} and $F^{(m)}$ in Eq.~\eqref{eq:curvature_matrix2} show that derivatives with respect to $d$ all vanish; in other words, radial or longitudinal derivatives vanish, and the geometry and topology are confined to a sphere in the parameter space, a consequence of the rotational covariance of ${\cal H}_\chi$. 
Indeed, from  Eq.~\eqref{eq:metric_matrix2} we can easily calculate the line element
\begin{align}\label{eq:line_element}
d s_m^2 &= g^{(m)}_{\theta \theta} \, d\theta^2 + g^{(m)}_{\phi\phi}\, d\phi^2 \nonumber \\
&=C_m\left(d \theta^2+\sin ^2 \theta \, d \phi^2\right) , 
\end{align}
which is the line element on a sphere $S^2$ of radius $R_m = \sqrt{C_m}$ and surface area $A_m=4\pi C_m$. The radius of this sphere depends on the band; for $S=1$, the geostrophic band with $C_0=1$ has a larger metric scale than the Poincar\'e bands which have $C_{\pm 1}=1/2$. 

\item In the pseudospin language $C_m$ has a simple physical interpretation in terms of transverse spin fluctuations. To see this, return to the pseudospin-$S$ Hamiltonian ${\cal H}_\chi = -\mathbf{d} \cdot \mathbf{S}$; the eigenstate $|S,m\rangle$ aligned with $\hat{\mathbf{d}}$ has
\begin{equation}
\langle\mathbf{S}\rangle=m \hat{\mathbf{d}},
\end{equation}
and the total spin length is fixed by
\begin{equation}
\left\langle\mathbf{S}^2\right\rangle=S(S+1) ,
\end{equation}
so the transverse spin fluctuations satisfy
\begin{align}
\left\langle\mathbf{S}_{\perp}^2\right\rangle & =\left\langle\mathbf{S}^2\right\rangle-\left\langle(\mathbf{S} \cdot \hat{\mathbf{d}})^2\right\rangle \nonumber \\
 &=S(S+1)-m^2=2 C_m . 
\end{align}
Therefore $2 C_m$ measures how much spin variance is available transverse to $\hat{\mathbf{d}}$. This is exactly what the FS metric measures: the sensitivity of state changes under an infinitesimal rotation of $\hat{\mathbf{d}}$, i.e., under transverse perturbations.

\item Both the FS metric \eqref{eq:metric_tensor} and the Berry curvature \eqref{eq:curvature1} inherit the $1/d^2$ ``Berry monopole'' from the QGT, suggesting that both may be integrated over a spherical surface to obtain constants. To see this for the FS metric, we introduce the area 2-form (the Riemannian volume form) \cite{nakahara2018geometry}
\begin{equation}
\mathrm{d} A_m \coloneq \sqrt{\mathrm{Det}\, g^{(m)}}\, d \theta \wedge d \phi .
\end{equation}
From Eq.~\eqref{eq:metric_matrix2} (we only include the upper left block in calculating the determinant)
\begin{equation}
\sqrt{\mathrm{Det}\, g^{(m)}} =C_m \sin \theta ,
\end{equation}
so the metric-induced area 2 -form is
\begin{equation}
\mathrm{d} A_m=C_m \sin \theta \, d \theta \wedge d \phi .
\end{equation}
Integrating over the sphere gives the total FS area $A_m$ for the $m$-th band [i.e. the Riemannian area of $\left(S^2, g^{(m)}\right)$ ],
\begin{align}
A_m &=\int_{S^2} \mathrm{~d} A_m \nonumber \\
&=C_m \int_0^\pi \sin \theta d \theta \int_0^{2 \pi} d \phi=4 \pi C_m ,
\end{align}
confirming our previous identification of the radius of the sphere as $R_m=\sqrt{C_m}$. From the Gauss-Bonnet theorem, the Euler characteristic of a manifold $M$ is \cite{nakahara2018geometry}
\begin{equation}
\chi_m(M)=\frac{1}{2 \pi} \int_M K_m d A_m,
\end{equation}
where $K_m$ is the Gaussian curvature of the metric $g^{(m)}$; for our sphere, $K_m=1/R_m^2 = 1/C_m$, and we have 
\begin{align}
\chi_m\left(S^2\right) & =\frac{1}{2 \pi} \int_0^{2 \pi} \int_0^\pi\left(\frac{1}{C_m}\right)\left(C_m \sin \theta\right) d \theta d \phi \nonumber\\ 
&= 2, 
\end{align}
so the Euler characteristic for all three bands is $\chi_m=2$. On the other hand, the Berry curvature is naturally a 2-form (see Eq.~\eqref{eq:Berry_twoform}); 
on $S^2$ with coordinates $(\theta, \phi)$, we have
\begin{equation}
\mathcal{F}_m=F_{\theta \phi}^{(m)} d \theta \wedge d \phi=m \sin \theta\, d \theta \wedge d \phi .
\end{equation}
Integrating over the sphere $S^2$,
\begin{equation}
\int_{S^2} \mathcal{F}_m=m \int_0^\pi \sin \theta\, d \theta \int_0^{2 \pi} d \phi=4 \pi m ,
\end{equation}
we obtain the quantized first Chern number
\begin{equation}
\mathcal{C}_m=\frac{1}{2 \pi} \int_{S^2} \mathcal{F}_m=\frac{1}{2 \pi}(4 \pi m)=2 m .
\end{equation}

\end{itemize}

To summarize, two independent pieces of band geometry and topology are controlled by two different invariants: the Chern number ${\cal C}_m = 2m$, which controls the topology, and the FS metric scale  $C_m=[S(S+1)-m^2]/2$, which controls the geometry and transverse spin fluctuations. For the specific case of the RSWE (with $S$=1), we have two Poincar\'e bands with non-trivial topology (Chern number ${\cal C}_\pm = \pm 2$) and Riemannian area $A_{\pm}= \pi$, and one geostrophic band with trivial topology (Chern number ${\cal C}_0 = 0$) and Riemannian area $A_0 = 4\pi$. All three bands have the same Euler characteristic $\chi_m=2$. Our results for the FS metric and the Berry curvature are conveniently summarized in Table~\ref{tab:QGT-metric-curv}.

\begin{table*}[t]
  \centering
  \small
  \setlength{\tabcolsep}{14pt}
  \renewcommand{\arraystretch}{1.3}
  \begin{tabular}{c c c}
    \hline
    \ &
    $g^{(m)}_{ij}$ (Fubini-Study metric) &
    $F^{(m)}_{ij}$ (Berry curvature)
    \\
    \hline\hline
    \shortstack{Definition \\ \ }
    &
                \shortstack{ 
                $g^{(m)}_{ij} = \text{Re}\,Q^{(m)}_{ij}$}
    &
    \shortstack{
                $F^{(m)}_{ij} = 2\,\text{Im} \,Q^{(m)}_{ij}$}
    \\  
    \hline
    \shortstack{Properties\\ [2pt] \\ [2pt]}
    &
    \shortstack{Real, symmetric \\[2pt]
                Positive semidefinite \\[2pt]
                Transverse: $g^{(m)}_{ij} d_j = 0$}
    &
    \shortstack{Real, antisymmetric \\[2pt]
                Gauge invariant \\[2pt]
                Monopole charge $2m$ at $\mathbf d=0$}
    \\[4pt]
    \hline  
    \shortstack{Cartesian form \\[2pt]
    ($i,j\in\{k_x,k_y,f\}$)}
    &
    $\displaystyle
      g^{(m)}_{ij}(\mathbf d)
      =
      C_m\left(
        \frac{\delta_{ij}}{d^2}
        - \frac{d_i d_j}{d^4}
      \right)
      $
    &
    $\displaystyle
      F^{(m)}_{ij}(\mathbf d)
      =
      m\,\frac{\epsilon_{ijk}\,d_k}{d^3}
      $
    \\[4pt]  
    \hline
    \shortstack{Spherical form \\[2pt]
    on $S^2$ ($\theta,\phi$)}
    &
    \shortstack{%
      $g^{(m)}_{\theta\theta} = C_m,\quad
       g^{(m)}_{\phi\phi} = C_m \sin^2\theta$ \\[2pt]
      $g^{(m)}_{\theta\phi} = g^{(m)}_{\phi\theta} = 0$}
    &
    \shortstack{%
      $F^{(m)}_{\theta\phi} = m \sin\theta,\quad
       F^{(m)}_{\phi\theta} = -\,m \sin\theta$ \\[2pt]
      $F^{(m)}_{\theta\theta} = F^{(m)}_{\phi\phi} = 0$}
    \\[4pt]
    \hline
  \end{tabular}
  \caption{Fubini-Study metric $g^{(m)}_{ij}$ and Berry curvature $F^{(m)}_{ij}$ for the
  pseudospin-1 RSWE bands labeled by $m\in\{-1,0,+1\}$. Cartesian indices $i,j$ run over
  $\{k_x,k_y,f\}$ with $\mathbf d=(k_x,k_y,f)$ and $d=|\mathbf d|$. The full quantum
  geometric tensor is $Q^{(m)}_{ij} = g^{(m)}_{ij} + \tfrac{i}{2} F^{(m)}_{ij}$. The
  band-dependent constants $C_m$ set the overall metric scale ($C_{+1}=C_{-1}=\tfrac12$, $C_0=1$). In Cartesian form, $g^{(m)}$ and $F^{(m)}$ are
  transverse to $\mathbf d$, reflecting the null direction along $\hat{\mathbf d}$.
  On the direction sphere $S^2$ of fixed $d$ (angles $\theta,\phi$), the metric components
  $g^{(m)}_{\theta\theta}, g^{(m)}_{\phi\phi}$ describe a round sphere of radius
  $\sqrt{C_m}$, while $F^{(m)}_{\theta\phi}$ describes a monopole Berry curvature
  with total flux $4\pi m$ and Chern number $\mathcal C_m = 2m$.}
  \label{tab:QGT-metric-curv}
\end{table*}

\section{Measuring components of the QGT}\label{sec:measurement}
Having developed the formalism for the quantum geometric tensor, how do we actually put it to use, and how can we measure it? As shown by Ozawa and Goldman~\cite{ozawa2018extracting,ozawa2019probing}, parametric driving of the control parameters can be used to measure components of the QGT \cite{tan2019experimental}---what one might call ``geometric spectroscopy''.  In geophysical flows, the nominal control parameters $(k_x, k_y, f)$  are not independently or sinusoidally tunable: the wavevector $\bf k$
 labels a wavepacket rather than an external knob, and the Coriolis parameter 
$f$ varies only slowly with latitude, not on experimental timescales. Natural forcing is broadband and inhomogeneous, while boundaries, stratification, dissipation, and turbulence all violate the spatial uniformity assumed by parametric driving. Consequently, such protocols are best viewed as surrogates for controlled laboratory tanks or numerical experiments, not realistic geophysical conditions.

To extract the QGT for RSWE using parametric driving, we start with $\cal H={\cal H}_\text{SWE}(\lambda)$ to be a  parameter-dependent
Hamiltonian with  ${\cal H}(\lambda)\ket{n(\lambda)}=\omega_n(\lambda)\ket{n(\lambda)}$,
$\braket{m|n}=\delta_{mn}$. For the RSWE, the parameter $\lambda_{\mu}=(k_x, k_y, f)$. We will treat \emph{small} sinusoidal modulations of
\emph{two} adiabatic parameters, $\lambda_\mu$ and $\lambda_\nu$, both at the same
drive frequency $\omega$ with a relative phase $\phi$:
\begin{align}
&\lambda_\mu(t)=\lambda^0_{\mu}+A_\mu\cos(\omega t),\\
&\lambda_\nu(t)=\lambda^0_{\nu}+A_\nu\cos(\omega t+\phi).   
\end{align}
Expanding to linear order in the amplitudes we obtain  ${\cal H}(\lambda)={\cal H}(\lambda^0)+V(t)$ with,
\begin{align}
\resizebox{\columnwidth}{!}{$
V(t)=\left(A_\mu\cos\omega t\,\partial_\mu{\cal H}+ A_\nu\cos(\omega t\, +\phi)\,\partial_\nu{\cal H}\right)\big|_{\lambda=\lambda^0}.
$}
\end{align}
In the above expression, we have defined $\partial_\mu=\partial/\partial\lambda_\mu$. 
Assuming the initial state of the system to be in band $n$, we apply \emph{Fermi's golden rule} \cite{landau2013quantum} and obtain the total rate for transitions $n\to m$ ($m\neq n$) at frequency $\omega$:
\begin{align}
&\Gamma^{(\mu\nu)}_{n_\rightarrow m}(\omega,\phi)
=\frac{\pi A_\mu A_\nu}{2}\nonumber\\\,
&\times\mathrm{Re}\!\Big[e^{i\phi}\,\langle n|\partial_\mu{\cal H}|m\rangle
\langle m|\partial_\nu{\cal H}|n\rangle\Big]\,
\delta(\omega-|\omega_{mn}|),
\end{align}
where we have defined $\omega_{mn}=\omega_m-\omega_n$. 
We now define an integrated transition. This overall factor could be a prefactor in the parametric drive and is necessary to match the exact form of the QGT in the observable:
\begin{align}
\mathcal I^{(n)}_{\mu\nu}(\phi)
=\sum_{m\neq n}\int_0^\infty \frac{\Gamma^{(\mu\nu)}_{n\rightarrow m}(\omega,\phi)}{\omega^2}\,d\omega.
\end{align}
The above expression can be written as
\begin{align}
\mathcal I^{(n)}_{\mu\nu}(\phi)& =\frac{\pi A_\mu A_\nu}{2}\, \nonumber \\
&\times \mathrm{Re}\!\left[e^{i\phi}\sum_{m\neq n}
\frac{\langle n|\partial_\mu{\cal H}|m\rangle
\langle m|\partial_\nu{\cal H}|n\rangle}{\omega_{mn}^2}\right] .
\end{align}
Using the Hellmann-Feynman identity
$\langle m|\partial_\mu{\cal H}|n\rangle=(\omega_n-\omega_m)\langle m|\partial_\mu n\rangle$, we can rewrite the above expression as
\begin{align}
\mathcal I^{(n)}_{\mu\nu}(\phi)=\frac{\pi A_\mu A_\nu}{2}\,
\mathrm{Re}\!\left[e^{i\phi}\,
\langle \partial_\mu n|(1-|n\rangle\langle n|)|\partial_\nu n\rangle\right].
\end{align}
Writing the quantum geometric tensor $Q^{(n)}_{\mu\nu}=g^{(n)}_{\mu\nu}+\tfrac{i}{2}F^{(n)}_{\mu\nu}$,
\begin{equation}
\mathcal I^{(n)}_{\mu\nu}(\phi)
=\frac{\pi}{2}\,A_\mu A_\nu\,
\Big[g^{(n)}_{\mu\nu}\cos\phi+\tfrac12 F^{(n)}_{\mu\nu}\sin\phi\Big] ,
\end{equation}
for the case when $\mu\neq \nu$. For the case when $\cal H=\cal H_\text{SWE}$ the $g^n_{\mu\nu}$ and $F^n_{\mu\nu}$ are given in Table \ref{tab:QGT-metric-curv}.

\section{Discussion and Conclusion}\label{sec:summary}

In this work we developed a quantum-geometric description of the rotating shallow water equations (RSWE) on the $f$-plane by computing the full quantum geometric tensor (QGT) for all three linear wave bands. Building on the now-standard observation that the linearized RSWE can be cast as a Hermitian three-band eigenvalue problem, we went beyond a purely topological characterization and extracted the gauge-invariant \emph{geometric} data encoded in the real part of the QGT (the Fubini--Study metric) together with the familiar \emph{topological} data encoded in the imaginary part (the Berry curvature). In the pseudospin formulation, rotational covariance strongly constrains the answer: the quantum metric is transverse to the radial direction in parameter space and induces a round metric on the direction sphere, while the Berry curvature takes the form of an effective monopole field whose flux gives the Chern number of the Poincar\'e bands. A key advantage of this geometric viewpoint is that it separates what is robustly quantized (integrated curvature) from what is not (the overall metric scale), thereby providing a systematic language for band-to-band comparisons and for diagnosing which physical effects are controlled by topology versus geometry.

A central conceptual message is that the RSWE inherit the same ``Riemannian + symplectic'' structure that appears in multiband quantum systems: $(g_{ij},F_{ij})$ together define, respectively, a notion of distance (state distinguishability) and a notion of geometric phase (adiabatic holonomy) on the manifold of control parameters. For the RSWE, this is not merely formal. The Berry curvature already has a clear wave-mechanical signature through anomalous transport of rays (geometric drifts), as emphasized in prior geophysical work \cite{perez2021manifestation,venaille2023ray}; the additional content of the present paper is that the metric supplies independent and, in principle, measurable information about the same bands. Operationally, the metric can be accessed through ``geometric spectroscopy'' protocols in which weak, time-periodic modulations of control parameters generate interband transitions whose integrated rates isolate specific components of $g_{ij}$ (and, with phase control, $F_{ij}$). This provides a potential route to extracting geometry in an analogue setting (of rotating tanks that can mimic RSWE) using controlled forcing, in close analogy with measurement proposals that have been developed for Bloch bands in condensed matter and synthetic-matter platforms \cite{ozawa2018extracting, ozawa2019probing, tran2017probing, tan2019experimental}.

There are several possible future applications of the QGT to geophysics. First, it will be important to embed the QGT into a systematic wave-packet theory for slowly varying backgrounds \cite{littlejohn1986semiclassical,sundaram1999wave,xiao2010berry,liang2017wave,lapa2019semiclassical}, where the Berry curvature controls leading post-eikonal corrections to ray trajectories. At the same time, the quantum metric is expected to enter at the same order as diffractive (paraxial) spreading of the envelope \cite{buhler2014waves,perez2021manifestation,venaille2023ray}. This direction would produce a unified "geometric optics + geometry" framework in which drift and diffraction are governed by intrinsic band data rather than ad hoc closure schemes. Second, the present $f$-plane results should be generalized to spatially varying Coriolis parameter and stratification (e.g., $\beta$-plane settings and smoothly varying profiles), where one expects a competition between local band geometry and gradients that can induce mode conversion. Third, it will be valuable to connect these geometric diagnostics to data analysis and observables in realistic flows: for example, identifying signatures of nontrivial band geometry in forced-wave experiments, in laboratory tanks, or in numerical simulations, and clarifying which aspects of the metric are robust to weak dissipation and weak nonlinearity. In these ways, quantum geometry can become a practical tool for geophysical wave physics, complementing and extending earlier topological perspectives.

\section{Acknowledgments}
	
We gratefully acknowledge discussions with Kinjal Dasbiswas, Paul Goldbart and Brad Marston. This work was initiated in part at the Aspen Center for Physics, which is supported by National Science Foundation grant PHY-2210452. SG is supported by NSF CAREER Grant No. DMR-1944967. ATD is supported by a grant from the Simons Foundation International [SFI-PD-Pivot Fellow-00008616, ATD].

\appendix

\section{Overview of the quantum geometric tensor} \label{app:overview}

To make this work relatively self-contained, in this appendix, we draw together some useful definitions and properties of the QGT. Several excellent reviews of the QGT and its applications have recently appeared in the literature, and we refer the reader to Refs.~\cite{cheng2013quantum,verma2025quantum,gao2025quantum,yu2024quantum} for additional background material. 

\textit{Definition and basic properties.}  
Let $\bm{\lambda}=(\lambda^1,\ldots,\lambda^N)$ denote a set of $N$ real control parameters and let $|u_m(\bm{\lambda})\rangle$ be a \emph{normalized} eigenvector of the $m$-th nondegenerate band  of a Hermitian Hamiltonian ${\cal H}(\bm{\lambda})$. The \emph{quantum geometric tensor} (QGT) is the gauge-invariant, complex Hermitian tensor is given by \cite{provost1980riemannian,berry1989quantum}
\begin{align}
Q_{ij}^{(m)}(\bm{\lambda}) &\coloneq \big\langle \partial_i u_m \,\big|\, \bigl( \mathds{1}_N - |u_m\rangle\langle u_m| \bigr)\,\big|\, \partial_j u_m \big\rangle \nonumber \\
&= g_{ij}^{(m)} + \frac{i}{2} F_{ij}^{(m)},
\label{eq:QGT_def}
\end{align}
where $\partial_i \coloneq \partial/\partial\lambda^i$ and where we have explicitly separated $Q_{ij}^{(m)}$ into its real, symmetric part $g_{ij}^{(m)} = \mathrm{Re}\ Q_{ij}^{(m)}$ and imaginary, antisymmetric part $F_{ij}^{(m)} = 2\, \mathrm{Im}\ Q_{ij}^{(m)}$ [where the factor of 2 and the plus sign in Eq.~\eqref{eq:QGT_def} are conventions]. The real part $g_{ij}^{(m)}$ serves as a metric [the Fubini-Study (FS) or quantum metric] on the parameter space \cite{fubini1904sulle,study1905kurzeste} and equips it with a \emph{Riemannian} structure, such that a line element for the $m$-th band is given by
\begin{equation}
    ds^2_m \coloneq g_{ij}^{(m)}\, d\lambda^i\, d\lambda^j.
\end{equation}
The imaginary part $F_{ij}^{(m)}$ can be expressed in terms of the \emph{Berry connection} $A_j^{(m)} = i\langle u_m|\partial_j u_m\rangle$ as
\begin{equation}
F_{ij}^{(m)} = \partial_i A_j^{(m)} - \partial_j A_i^{(m)},
\label{eq:Berry_from_A}
\end{equation}
and the corresponding \emph{Berry phase} on a loop $\Gamma$ is $\gamma_\Gamma=\oint_\Gamma A^{(m)}$. Equivalently, $F_{ij}^{(m)}$
defines a \emph{closed 2-form}
\begin{equation}
F^{(m)} = \frac{1}{2}\,F_{ij}^{(m)}\,d\lambda^i\wedge d\lambda^j,
\label{eq:curvature_form}
\end{equation}
which is the curvature of the $U(1)$ Berry connection. On a closed, oriented $2$-manifold $\Sigma$,
\begin{equation}
{\cal C}_m=\; \frac{1}{2\pi}\int_{\Sigma} F^{(m)} \;\in\; \mathbb{Z}
\label{eq:Chern_number}
\end{equation}
is the \textit{first Chern number} (which is topologically quantized).

\textit{Projector formulation.} For some applications, it is more convenient to work in terms of the band projectors $P_m(\bm{\lambda})\coloneq |u_m\rangle\langle u_m|$, which are gauge invariant. In terms of the projectors,
\begin{align}\label{eq:projector}
Q_{ij}^{(m)} (\bm{\lambda}) &= \mathrm{Tr} \left[ (\partial_i P_m)(1- P_m) (\partial_j P_m) \right] \nonumber\\
&=\mathrm{Tr} \left[ P_m (\partial_i P_m) (\partial_j P_m) \right] . 
\end{align}
Additional identities and properties of the projector calculus may be found in Refs.~\cite{graf2021berry,mitscherling2025gauge,avdoshkin2025multistate}.

\textit{Positivity.} The QGT is a \textit{Gram matrix} formed from the projected tangent vectors $|v_i\rangle=(1-P_m)|\partial_i u_m\rangle$, 
\begin{align}
    Q_{ij}^{(m)} &= \langle v_i|v_j\rangle = \langle \partial_iu_m|(1-P_m)(1-P_m)|\partial_j u_m\rangle \nonumber \\
    &= \langle \partial_i u_m |(1 - P_m)| \partial_j u_m \rangle,
\end{align}
where we have used $P_m^2=P_m$ ($P_m$ is an \textit{idempotent} operator). The QGT is therefore has a positive semi-definite Hermitian form,
\begin{align}
\sum_{i,j} a_i^* Q_{ij}^{(m)} a_j 
 &= \sum_{i,j} a_i^*\langle v_i|v_j\rangle a_j \nonumber \\
 &=\left(\sum_i a_i | v_i\rangle \right)^\dagger \left(\sum_i a_i | v_i\rangle \right)  
\ge 0, 
\label{eq:PSD}
\end{align}
for all complex vectors $\boldsymbol a$. As a result the eigenvalues of $Q_{ij}^{(m)}$ are real and non-negative, and therefore $\text{Tr}\ Q^{(m)} \ge 0$ and $\text{Det}\ Q^{(m)} \ge 0$. 

\textit{Spectral representation.}
Using the Hellman-Feynman identity
\begin{equation}
    \langle u_n|\partial_i {\cal H} | u_m\rangle = (\omega_m - \omega_n) \langle u_n|\partial_i u_m\rangle
\end{equation} 
for a nondegenerate eigenvector $|u_m\rangle$ of ${\cal H}$ with eigenvalue $\omega_m$, we obtain a spectral representation for the QGT,
\begin{equation}
Q_{ij}^{(m)} \;=\; \sum_{n\neq m} 
\frac{\langle u_m|\partial_i {\cal H}|u_n\rangle\,\langle u_n|\partial_j {\cal H}|u_m\rangle}{(\omega_m - \omega_n)^2},
\label{eq:spectral}
\end{equation}
where $n\neq m$ runs over the other bands. The FS metric and the Berry curvature can be obtained by taking the real and imaginary parts of this expression. This expression is basis and gauge-independent and particularly convenient when $\partial_i {\cal H}$ has a simple structure.

\textit{Inequalities linking $g$ and $F$.}
For a two dimensional parameter space (say $\lambda^1,\lambda^2$) for a single band,
\begin{equation}
4\,\mathrm{Det}\ g^{(m)} \;\ge\; \left(F_{12}^{(m)}\right)^{2}.
\label{eq:det_inequality}
\end{equation}
This follows from the positivity of the $2\times 2$ Hermitian matrix $Q_{ij}^{(m)}=g_{ij}+\tfrac{i}{2}F_{ij}$ \cite{kolodrubetz2017geometry,ozawa2021relations, mera2021kahler}. A useful corollary is
\begin{equation}
\mathrm{Tr}\,g \;\ge\; |F_{12}^{(m)}|,
\label{eq:trace_bound}
\end{equation}
obtained by the arithmetic/geometric bound $\mathrm{Tr}\,g \ge 2\sqrt{\mathrm{Det}\ g}$ and \eqref{eq:det_inequality}. Equality in \eqref{eq:det_inequality} holds when the state manifold inherits a K\"ahler structure from the projective Hilbert space (e.g., two-level systems mapped to a Bloch sphere). In those cases, $g$ and $F$ are the real and imaginary parts of a single K\"ahler form \cite{mera2021kahler,nakahara2018geometry,bengtsson2017geometry}.

\bibliography{qgt_rswe}

\end{document}